# Hindered rotation of $H_2$ adsorbed interstitially in nanotube bundles.

M.K. Kostov [†], and H. Cheng

Computational Modeling Center and Corporate Science and Technology Center,

Air Products and Chemicals, Inc. 7201 Hamilton Boulevard, Allentown, PA 18195-1501

R.M. Herman, and M.W. Cole

Department of Physics, Penn State University, University Park, Pennsylvania 16802

J.C. Lewis

Department of Physics & Physical Oceanography, Memorial University of Newfoundland,

St. John's, NF Canada A1B 3X7.

## ABSTRACT

A theoretical study on the rotational dynamics of $H_2$ molecules trapped in the interstitial channels (ICs) of a carbon nanotube bundle is presented. The potential used in this study is modeled as a sum of atom-atom (C-H) van der Waals interactions and electrostatic interactions of the molecule with the surrounding nanotubes. The rotational energy spectra is calculated using a product wave function, where the coupling between translational and rotational modes is treated in a mean-field manner. Molecular dynamics (MD) simulation study was performed for estimating the hydrogen rotational barrier. Both theoretical calculations and simulation results reveal the existence of a large rotational barrier (~ 40 meV). The consequences of this rotational barrier for the rotational energy levels are worked out in detail.

## I. Introduction

Recent progress in material preparation has led to the production of quite regular porous media having characteristic widths of the order of few nanometers[1]. These include various assemblages of carbon nanotube materials, including single-wall carbon nanotubes (SWNT) or multi-wall nanotubes. Recent experiments have proved that one can create ordered arrays of nanotubes[2,3], which form a closed – packed bundle similar to strands of bucatini held tightly in one's hand. Such a bundle of nanotubes provides an environment in which small atoms can fit well in the interstitial channels (ICs) between closed – packed nanotubes [2,4,5]. Due to the remarkable geometry of the nanotubes( very small diameter ~ 14Å, relatively long length ~ 10μm), the IC's environment has a large number of neighboring C atoms at an optimal distance for small atoms and molecules, for which the interstitial binding energy is larger than is found in any other known environment[16]. Because of this strong binding, there has been much interest in the properties of small molecules confined to the interstitial regions within a nanotube bundle.

It is known that the physisorption interactions of nonspherical molecules with surfaces can alter their rotational behavior [6]. One expects that the extreme anisotropy of the ICs will yield an orientation-dependent interaction of nonspherical molecules confined there, which will hinder their rotational motion. The case of hydrogen is a particularly appealing system to study, since it has low mass, is well defined as a quantum, nearly rigid rotor, and its adsorption in ICs is relevant to both gas storage and isotope separation[4]. In this paper, we present a simplified quantum-mechanical model of the rotational dynamics of $H_2$ molecules trapped in the ICs of a carbon nanotube bundle, assuming perfectly conducting nanotube walls. We then compute the hindered rotational spectra and compare the results with recent experiments.

The rotational motion of $H_2$ in free space, is to good approximation, that of a simple rigid rotor:

$$E_J = B\,J\,(J+1) \quad , \tag{1}$$

where $B = 7.35$ meV is the rotational constant of $H_2$ in its ground vibrational state and $J$ is the angular momentum quantum number[7]. This simple expression is a consequence of the isotropy of free space and the neglect of rotational-vibrational coupling. External fields, such as those existing at the surface of a solid, destroy spatial isotropy and lift the degeneracy in the azimuthal quantum number $M$ in ways determined by the symmetry and intensity of the field at the $H_2$. The resulting energy levels thus provide information about the environment. Techniques which have measured such effects include infrared absorption, Raman scattering and neutron scattering[5,8]. We treat the interaction between the molecule and nearby nanotubes with a semiempirical method analogous to that used previously to treat adsorption on graphite surfaces[9,10]. The method employs an empirical isotropic Lennard-Jones (LJ) pair potential between each atom of the molecule and each carbon atom of the surrounding surface. The potential also includes an orientation-dependent quadrupole interaction between the $H_2$ molecule and the local electrostatic field provided by the carbon. Such an approximation was used recently in calculations of $N_2$ on graphite[11–13]. These various papers disagree about the sign of this term, which is an important component of the total energy and especially the orientation dependence. Our approach differs somewhat from that of the earlier work because it incorporates an *ab initio* calculation of the electrostatic field [17]. The rotational states were calculated using a product wave function consisting of a rotational part, multiplied by a spatial part, which describes the motion of the molecular center of mass (CM). The rotational term is written as an expansion in free-rotor basis functions and the interaction of translational and rotational modes is treated in a mean-field manner [9]. In addition, we have conducted fully atomistic molecular dynamics (MD) simulations to estimate the magnitude of hydrogen rotational barrier. Our theoretical calculations and simulation results indicate the existence of significantly hindered rotational motion of the $H_2$ molecules. Some preliminary neutron scattering data has been shown to be in semiquantitative agreement with our findings [8].

Transitions among the perturbed energy levels of the adsorbed $H_2$ molecules are allowed in Raman scattering and in neutron scattering, as is true for free $H_2$ molecules. However, the quadrupole moment of the $H_2$ will induce a net dipole moment in the carbons of the nanotubes [15], and in consequence, optical dipole transitions associated with these induced dipoles are also possible. The intensities and selection rules for these transitions will be the subject of a later publication.

This paper is organized as follows. In Section II our model and results for the potential energy as a function of orientation are reported. Section III presents and discusses results for the energy spectrum which follows from this potential. Section IV describes an alternative semiempirical MD calculation of the rotational barrier, which employs consistent valence forcefield and includes relaxation of the neighboring carbon atoms. Section V provides a summary and discussion of this work.

## II. Rotation of interstitial $H_2$ in SWNT

## A. The potential function

In our model we do not take into account the internal interactions in the $H_2$ molecule, i.e. we consider the molecule to be rigid dumbbell. There exists experimental evidence consistent with that assumption in that the $H_2$ vibrational spectrum is shifted by only ~ 1 cm$^{-1}$ in the IC environment[5]. Furthermore, we ignore the mutual interactions between $H_2$ molecules as well as the effect of $H_2$ on the host lattice[*]. Hence, our model neglects all cooperative effects, which we believe should not have an important influence on the rotational and vibrational excitations of individual $H_2$ molecules. Another assumption is to neglect the corrugation of the potential along the axial direction of the ICs and to average angularly the potential.

The potential felt by a molecule is modeled as a sum of atom-atom (C-H) van der Waals interactions and the electrostatic interaction of the molecule with three surrounding nanotubes :

$$V(z_c, \Omega) = U_{el} + U_{LJ}(z_1) + U_{LJ}(z_2). \qquad (2)$$

Here, the $z_i$ are the perpendicular distances of the H atoms relative to the surface ; $z_c$ is the perpendicular distance from the molecular CM to the surface atoms, and $\Omega = (?, f)$ is the orientation of the molecule's internuclear axis with respect to the IC

---

[*] Recent zero temperature calculations have found that tubes dilate slightly in order to accommodate the $H_2$ [24].

axis(z axis) and an x axis directed toward a neighboring tube's center. The dispersion part of this potential ignores many-body effects [14], considers the C atoms to be smeared on the surface, and assumes the C atoms to be isotropic. We have employed a potential, $U^*_{LJ}$, previously developed for a single nanotube[16]. When averaged over the azimuthal and longitudinal coordinates, this potential takes the form :

$$U^*_{LJ}(z) = 3\pi n \varepsilon_{C-H} \sigma^2_{C-H} \left[ \frac{21}{32} \left( \frac{\sigma_{C-H}}{R} \right)^{10} f_{11}(x) M_{11}(x) - \left( \frac{\sigma_{C-H}}{R} \right)^4 f_5(x) M_5(x) \right], \quad (3)$$

where $z$ is the distance from H atom to the axis of a nanotube, $R$ is the nanotube radius, and $n = 0.38\ ?\ ?^2$ is the surface density of the C atoms. We adopted the values [9] $\varepsilon_{C-H} = 2.26$ meV, $\sigma_{C-H} = 2.76$ Å and defined $x = R/z$ and $f_l(x) = (R/z)^l$. The $M_l$ are integrals defined in Ref.[16]. The potential for an IC is obtained by summing $U^*_{LJ}$ over three adjacent nanotubes and azimuthally averaging the result.

The electrostatic portion of the potential in the ICs contains two contributions, denoted $U_1$ and $U_2$, which arise, respectively, from the interaction of the H$_2$ quadrupole moment with the local electrostatic field of the carbon nanotubes, and the interaction of the H$_2$ static multipole moments with the image charges induced on the surface. $U_1$ is given by:

$$U_1(z_c, \cos\theta) = -\frac{1}{6} \sum_{i,j} \Theta_{ij} \frac{\partial^2 \Phi(z_c)}{\partial x_i \partial x_j}, \quad (4)$$

in which the $\Theta_{ij}$ are the components of the quadrupole moment and $\Phi$ is the local electrostatic potential, which was calculated from first principles[17] for so-called "zig-zag" (18,0) carbon nanotubes[18] with radius R = 6.9 ? .

To evaluate $U_2$, we included the quadrupole moment and the hexadecapole moment of the $H_2$. This problem is quite complicated, so we introduce three simplifications. Due to the small size of the molecule, we consider its multipoles to be "point"-like. Next, to simplify the complex geometry of the IC, we represent the carbon nanotube walls with graphene planes which are tangent to the actual tubes. Thus, an $H_2$ molecule at the origin is located at the center of an equilateral triangle, formed by three graphene sheets. Finally, we assumed that these sheets are perfectly conducting, so that the dynamic responses are represented by full strength images of molecular charge fluctuations; image planes lie a distance $z_0$ above the graphene sheets. The screened electrostatic interaction of $H_2$ with a single plane can be expressed in terms of the $H_2$ quadrupole(T) and hexadecapole (F) moments in the following way [10]:

$$U^*_2(z_c, \cos x) = -\frac{3k\Theta^2}{8(2LI)^5}[3\cos^4(x) + 2\cos^2(x) + 3] - \frac{15k\Theta\Phi}{16(2LI)^7}[7\cos^6(x) + 5\cos^4(x) + 9\cos^2(x) - 5], \quad (5)$$

where ? is the angle of the molecular axis with respect to the surface normal, $LI = z_c - z_0$ is the location of the image with respect to the graphene surface, $k$ is the electrostatic constant $k = 9.10^9 \frac{Nm^2}{C^2}$. The values for the $H_2$ moments, taken from Ref.[20], are T = 0.49 ; F = 0.23 in atomic units. The least well-known parameter in our model potential is $z_0$. To obtain this parameter we constructed the holding potential of $H_2$ on graphene surface as the sum of Lennard-Jones type atom-atom (C - H) interactions and the electrostatic energies, calculated with Eq.(5). For simplicity we smear out the C atoms along the surface. Next, we compare our holding potential with the isotropic $H_2$/graphene potential at the equilibrium distance $z_{eq} = \sigma_{C-H_2}$, taking into account that $H_2$ is preferentially oriented flat against the surface at $z_{eq}$. Using[20] $\varepsilon_{C-H_2} = 42.8$ K and $\sigma_{C-H_2} = 2.97$ Å , the resulting well depth is 46.6 meV, which yields $z_0 = 1.36$ Å , a value consistent with previous *ab initio* and empirical estimates for this system[14].

In our problem for interstitially adsorbed $H_2$ we have much more complicated geometry than a single graphite plane. To obtain $U_2$ we sum up the contributions $U^*_2$ from all image charges. The locations of the image charges are determined by the boundary condition requiring that the potential be zero at each plane.

### B. Rotational levels

Next we study the splitting of the rotational levels due to the interaction potential $V(z_c, \Omega)$, taking into account the effect of the zero-point motion. In general, the holding potential (Eq.2) can be expressed in terms of spherical harmonics:

$$V(z_c, \Omega) = \sum_L \sum_{M=-L}^{L} V_{LM}(z_c) Y_L^M(\Omega), \qquad (6)$$

where the coefficient $V_{LM}$ depends on the position $z_c$ of the molecular CM ; $\Omega = (?, f)$ is the orientation of the molecular axis with respect to the axis of the IC ( note that $? = p/2 - ?$ ). Since the anisotropic part of the molecule-substrate interaction depends on $z_c$, the translational and rotational parts of $V(z_c, \Omega)$ do not decouple. To separate the variables approximately, we treat the interaction of translational and rotational modes in a mean-field manner, following Novaco and Wroblewski [9]. The total wave function, then, is approximated by the product of rotational and translational wave functions. The key assumption in this approach is to average the anisotropic interaction over the range of the zero-point motion of the molecule, while keeping the orientation, $\Omega$, fixed:

$$\tilde{V}(\Omega) = <0|V(z_c,\Omega)|0>. \qquad (7)$$

In order to perform the required averaging, we treat the CM vibrations of the molecule as modes of a simple harmonic oscillator ( the ground state wave function $|0>$ is of Gaussian type). Utilizing the model potential $V(z_c, \Omega)$ we find the *rms* deviation in the z-position of molecular CM to be $dz_{rms} = 0.225$ Å. Most of the terms in the anisotropic potential (Eq.6) vanish due to the high symmetry of the IC geometry. In our approach the leading terms in the effective anisotropic potential (6) are $V_{00}$, $V_{20} Y_2^0(\Omega)$ and $V_{40} Y_4^0(\Omega)$. Carrying out numerical computations for the dispersion and electrostatic portions of $V(z_c, \Omega)$ and zero-point averaging, we obtain the surface orientational potential as a function of $\cos^2 \theta$ (see Fig.1). The magnitude of the orientational potential barrier is quite large : $V_B = 6.7 B$, which implies a significantly hindered rotational behavior of the interstitially adsorbed $H_2$ molecules. Adopting the effective anisotropic potential from Fig. 1 and separating the variables in Schrödinger equation, we reach :

$$\frac{d}{dh}\left[(1-h^2)\frac{d\Xi}{dh}\right] + \left[\frac{E_{qf}}{B} - C - C_2 h^2 - C_4 h^4 - \frac{M^2\hbar^2}{2I(1-h^2)}\right]\Xi = 0 \ , \qquad (8)$$

where $\Xi$ is the rotational wave function; $h \equiv \cos\theta$ ; $C = -15.0$, $C_2 = -3.0$, and $C_4 = -3.7$. If $C_4$ is taken equal to zero in Eq.(8), then the differential equation reduces to that obtained by separation of variables in the Helmholtz wave equation in spheroidal oblate coordinates [21,22]. The boundary conditions for (8) are such that $\Xi$ is bounded and single valued at $\theta = \pm 1$. Applying these boundary conditions, the eigenvalue problem for Eq.(8) was solved numerically using a "shooting" procedure based on a fixed-stepsize Cash-Carp Runge-Kutta method. The procedure was tested, for $C_4 = 0$, against an existing computer program [23] and published tables of eigenvalues[21,22] of spheroidal wave functions for a wide range of values of $C_2$ and of energy levels.

It is interesting to note some specific features of Eq.(8) in the limit of high rotational barriers. For the low-lying rotational states, and high barrier, the rotational potential experienced by the molecule approximates to:

$$V(q) = -(|C_0| + |C_2| + |C_4|) + (|C_2| + 2|C_4|)q^2 \qquad (9)$$

Hence, for large $|C_2|$ and/or large $|C_4|$, the energy levels approximate to those of a harmonic oscillator, with rotational fine structure superimposed, but each energy level corresponds to two wave functions ? , one of which is even under ? ? ($p - ?$) and the other odd. The asymptotic behavior of solutions of Eq.(8) is thus qualitatively the same as the well-known behavior [21] of the oblate spheroidal angle functions, i.e. of solutions of Eq. (8) with $|C_2| > 0$ and $|C_4| = 0$. The argument of section 8.2.1 of ref. [21] can in fact be extended in a straightforward way to show that for large values of $(|C_2| + |C_4|)^{1/2}$ the energy levels are given by:

$$E_{JM} = -(|C_0| + |C_2| + |C_4|) + 2\frac{(M + 1 + N)}{\sqrt{|C_2| + |C_4|}} + O(1), \qquad (10)$$

where $N = J - |M|$ for $J - |M|$ even; and $N = J - |M| - 1$ for $J - |M|$ odd. Hence as the barrier height increases, pairs of energy levels with the same value of $M$ coalesce: the ($J=0$, $|M|=0$) state with the (1,0) state, the (1,1) states with the (2,1) states, and so forth. Because the energy of the free rotator increases quadratically with $J$, the coalescence of energy levels requires increasingly large barriers for the higher states, and in the present problem is of importance only for $J=0$ and $J=1$.

## III. Results

Results for the rotational transition energies are listed in Table 1, where they are compared to the deep-inelastic neutron scattering (DINS) experimental values of Narehood *et al.* [8] . In these recent experiments the rotational transitions and the mean kinetic energy of the $H_2$ (in the impulse approximation limit) were measured. The experimental spectra reveal the existence of an orientation-dependent interaction between the $H_2$ and the nanotubes. Our results are consistent with this experimental observation. The rotational transition energies from the ground state are significantly different from the free molecule's values (see Table 1), indicating that the molecule experiences a significant orientational potential. In contrast, for $H_2$ on grafoil no shifts of the rotational transitions have been reported .

The discrepancy between theoretical and experimental transition energies, involving $J = 1$ rotational state, is tentatively attributed to our approximate treatment of the rotational-translational coupling . To our knowledge, a proper treatment of this coupling has not, so far, been carried out even in the simpler case of adsorbed $H_2$ on graphite[9] . In quantum language, the nonseparability between the translational and rotational coordinates produces a mixing of the $M = 0$ and $M = \pm 1$ states of the $J = 1$ manifold. These levels should exhibit the dominant coupling effects, because of the small spacing between them without the mixing. The other rotational states will also mix with the $J = 1$ states, but due to the large spacing between them and $J = 1$ manifold, the coupling effects will be relatively small. In support of this, we note that our theoretical results are in very good agreement with the experimental data for energy transitions not involving the $J = 1$ state.

## IV. Determination of rotational barrier by molecular simulations

Here, using a fully atomistic description of SWNTs, we provide an alternative approach to estimate the magnitude of the orientational potential barrier for interstitially adsorbed hydrogen. Our computational technique employs classical MD simulations, where Newton's equations of motion are numerically integrated with a Verlet velocity integrator algorithm [25]. The MD simulations enable us to study the dynamical coupling between the hydrogen and the nanotubes. The potential used in this section differs because of both the relaxation of the tubes, due to the $H_2$ molecules, and the use of a somewhat different potential energy model. The dynamics and energy minimization calculations are implemented utilizing the consistent valence forcefield (CVFF)[26]. For the sake of simplicity, we limit our MD simulations to bundles comprised of (18,0) nanotubes ( diameter ~ 13.8 Å ). The nanotubes in the simulation box are placed at the corners of equilateral parallelogram ( two-dimensional rhombic lattice ) and the SWNTs are assumed to be aligned with their axes parallel to each other. To model the longitudinal interactions accurately, the unit cell selected for our computational investigation contains 10 primitive cells, totaling 720 carbon atoms. The optimized unit cell parameters are $a = b = 16.78$ Å , $c = 42.60$ Å ; a = ß = 90° , and ? = 120° , where the size of the gap between two adjacent tubes is about 3 Å . Repeating this unit cell leads to a trigonal two-dimensional lattice of nanotubes. Periodic boundary conditions were applied in all three directions. The simulation study was performed for ~ 0.46 wt % gravimetric hydrogen capacity ( 2 $H_2$ + 72 C in a primitive cell). At this dilution, one can ignore the mutual $H_2$ interactions.

To determine the rotational barrier, we have conducted an energy minimization of the SWNT lattice, while keeping the orientation of the $H_2$ molecules fixed. Two orientations of the molecule were considered: the parallel with respect to the IC's axis (Fig. 2) and the transverse orientation (i.e. that perpendicular to the IC's axis, as shown in Fig. 3). The rotational barrier is determined from the total energy difference in the two energy minimization procedures. Within the CVFF forcefield, the nonbonding H-H, C-C, and C-H interactions are represented by van der Waals interactions characterized by a (12,6) Lennard-Jones function. The interactions between $H_2$ molecules are well modeled by the potential of Diep and Johnson [27]. In this work, we used H-H potential parameters which yield binding energies in good agreement with Diep and Johnson's potential for " T " $(H_2)_2$ (44K) and "slipped-parallel" $(H_2)_2$ (38K) orientations of the $H_2$ dimer at center of mass separation 3.4 Å between the two molecules. The C-C Lennard-Jones parameters utilized in our simulation are those derived from the properties of a graphite crystal [28] $\varepsilon_{C-C}$=2.41 meV and $\sigma_{C-C}$ = 3.40 Å. To model the C-$H_2$ interaction we adopted the C-H parameters used for $H_2$ on graphite[9]. The proposed approach yields a significant barrier to hydrogen rotational motion: $V_B$ = 35 meV = 4.8$B$ (about 70% of the value obtained in Section II). We believe that this result is a lower limit for the rotational barrier for the following reasons. In this study we assume a purely $sp^2$ carbon bonding network. The curvature of the SWNT, coupled with further nanotube deformations, should necessarily create regions of mechanically distorted C atoms with bonding configurations approaching $sp^3$. The hydrogen sorption mechanism, however, differs dramatically in the case of $sp^3$ carbon hybridization, because a partial electron-transfer interaction between distorted carbon atoms in the SWNT wall and $H_2$ is possible[29]. Consequently, an increased C-H interaction will lead to larger orientational potential barrier to the hydrogen rotational motion.

## V. Summary

We have presented two complementary models of the hindered rotation of interstitial $H_2$ in a nanotube bundle, with implications for the rotational spectrum. Both models employ traditional approximations concerning the force laws, supplemented by results of *ab initio* calculations of the electrostatic field[17]. The models yield similar(order of magnitude) results, consistent with the experimental observation of a large rotational barrier. It is argued that the mixing of $M = 0$ and $M = \pm 1$ states of the $J = 1$ manifold can explain the discrepancy between our theoretical and experimental results. A proper treatment of the rotational-translational coupling is beyond the scope of this paper. We emphasize that all approximations made in our model introduce an error, but the qualitative trends in the orientational potential ought to be reliable.

We are grateful to Dragan Stojkovic, Paul Sokol, David Narehood, Keith Williams, and Peter Eklund for helpful discussions. This research has been supported by the Petroleum Research Fund of the American Chemical Society and the Army Research Office.

# References

† Corresponding author. Email: mkk143@psu.edu . Fax: +1–814–865–3604.

Permanent address: Department of Physics, Penn State University, University Park, Pennsylvania 16802

Figure captions.

Fig. 1. Surface orientational potential as a function of $cos^2 ?$ .

Fig. 2. The SWNT primitive cell, selected for the energy minimization simulation, in the case of parallel molecular orientation with respect to the IC axis.

Fig. 3. The SWNT primitive cell, selected for the energy minimization simulation, in the case of transverse molecular orientation with respect to the IC axis.

**TABLE 1** Transition energies ( meV ) predicted and measured (DINS data) for indicated transitions[8]. In brackets are given the free space transition energies.

| (J,M) → (J', M') | Predicted | Measured |
|---|---|---|
| (0,0) → (1,0) | 5.1 (14.7) | 1.2 |
| (0,0) → (1,1) | 24.6 (14.7) | 17.3 |
| (0,0) → (3,0) | 84.3 (88.2) | 83.2 |
| (1,0) → (4,0) | 139 (147) | 158 |
| (0,0) → (5,0) | 217 (221) | 217 |

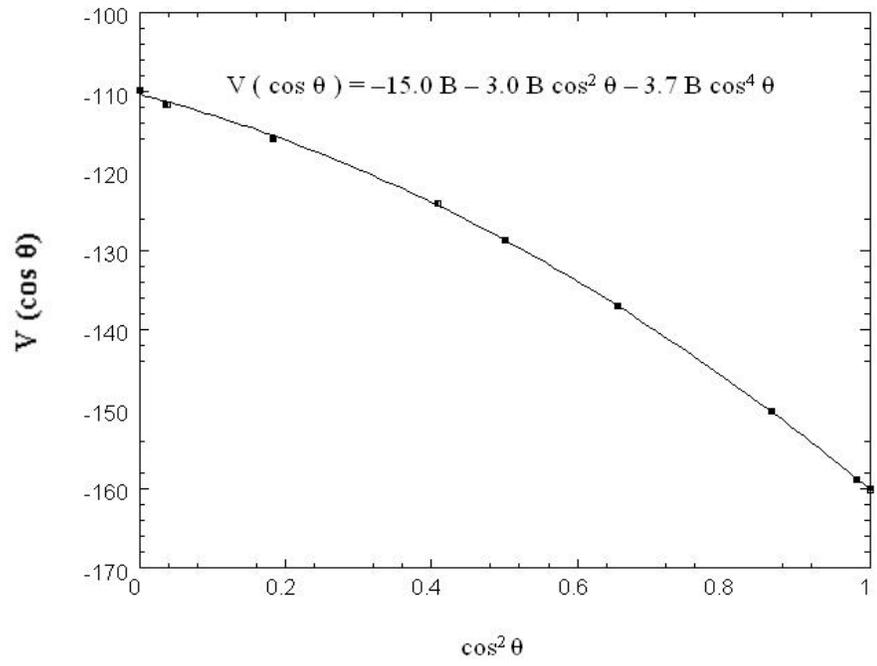

Fig. 1

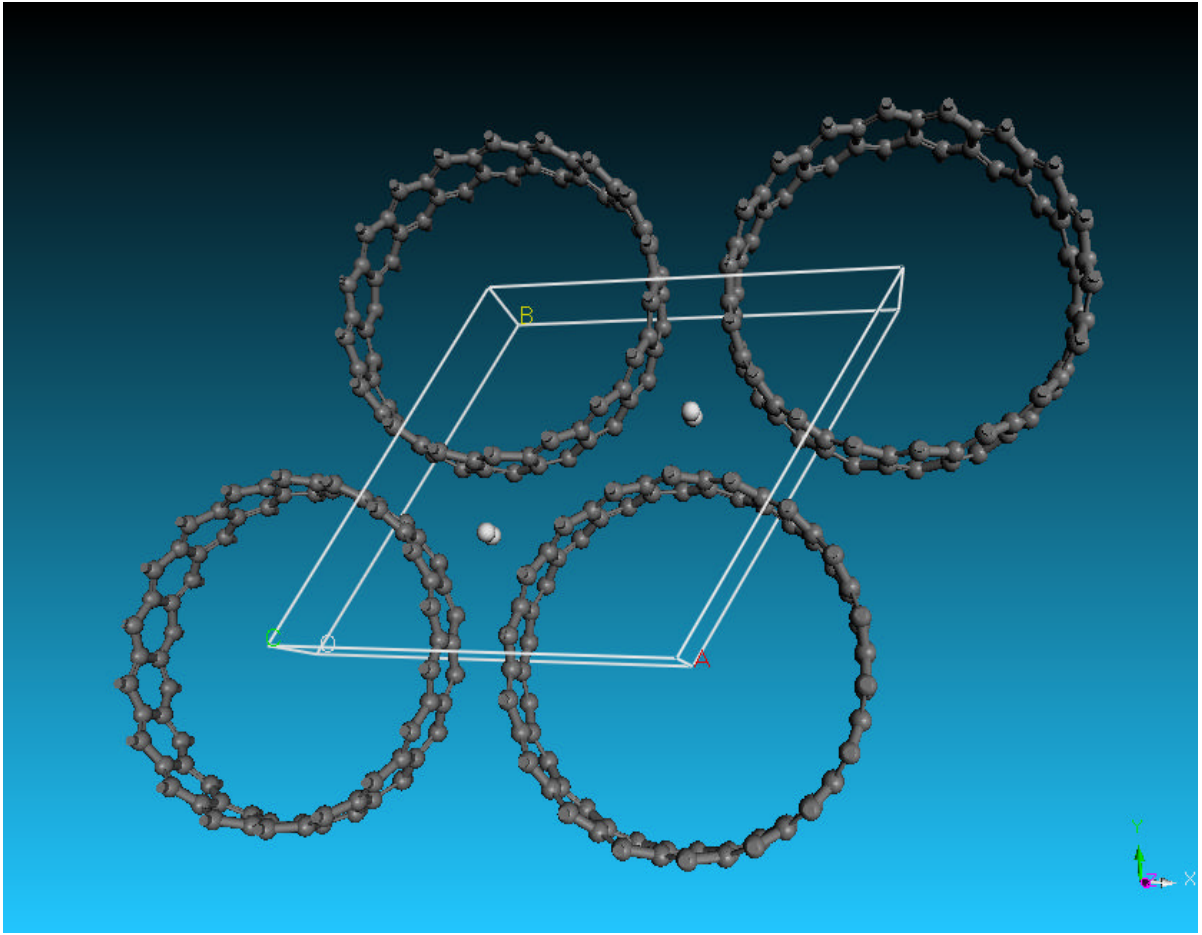

Fig. 2

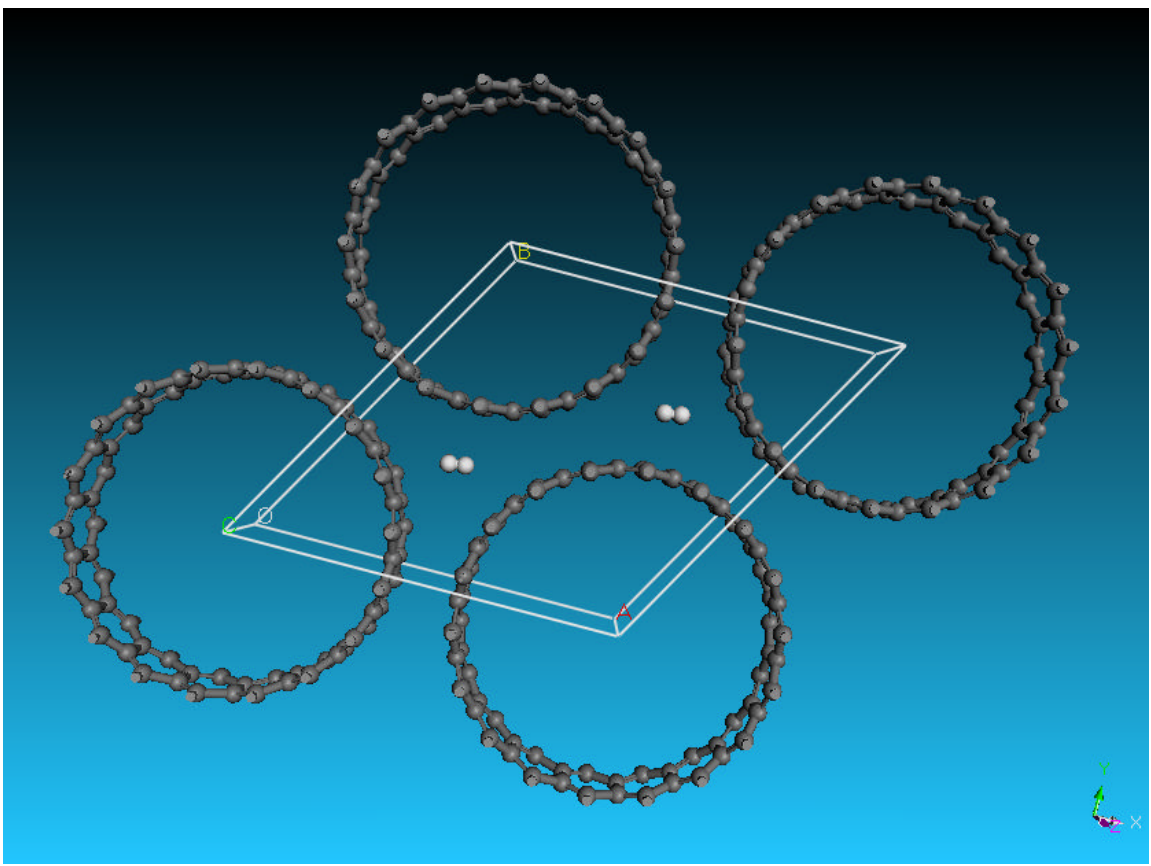

Fig. 3